# Induced magnetic moment at two-dimensional MXene/ferromagnetic interface evaluated by angle-dependent hard X-ray photoemission spectroscopy

Prabhat Kumar [1], Shunsuke Tsuda [2], Koichiro Yaji [2,3], and Shinji Isogami [1*]

1. Research Center for Magnetic and Spintronic Materials, National Institute for Materials Science (NIMS), Tsukuba 305-0047, Japan

2. Center for Basic Research on Materials, National Institute for Materials Science (NIMS), Tsukuba 305-0003, Japan

3. Unprecedented-scale Data Analytics Center, Tohoku University (UDAC), Tohoku University, Sendai 980-8578, Japan

*isogami.shinji@nims.go.jp

**Abstract**

Emergent ferromagnetism on the surface of recent two-dimensional (2D) MXene is investigated by X-ray magnetic circular dichroism (XMCD) and angle-dependent hard X-ray photoemission spectroscopy (HAXPES). Focusing on the $Cr_2N$ as one of the 2D-MXenes, the bilayers of $Cr_2N$/Co and $Cr_2N$/Pt are prepared by magnetron sputtering technique. XMCD reveals the induced magnetic moment of Cr in the $Cr_2N$/Co interface, while it is not observed in the $Cr_2N$/Pt interface at room temperature. To distinguish the possible origins of either the interlayer magnetic exchange coupling or the charge transfer at the interfaces, the additional controlled $Cr_2N$/Cu bilayer, whose work function of Cu is consistent with Co, is prepared. HAXPES spectra for the Cr $2p$ core level near the interface of $Cr_2N$/Cu is consistent with that of $Cr_2N$/Co, indicating that the induced magnetic moment of Cr observed by XMCD for the $Cr_2N$/Co can be attributed to the interlayer magnetic exchange coupling, rather than the charge transfer, which is a specific characteristics emerged at the interface with 2D-MXene.

# 1. Introduction

Highly efficient semiconductor and/or magnetic devices are essential for the development of a modern society in which humans are connected to all kinds of applications via the Internet. In order to realize these devices, the reduction of the bit cell size towards the existing technology node and the reduction of the power consumption have been desired. Two-dimensional (2D) materials are considered as one of the candidates, not only because of their structural advantages, but also because of various emergent phenomena at the interfaces of 2D heterojunctions, such as interlayer exchange coupling,[1] magnetoelectric effect,[2] and proximity effect.[3,4]

One of the most attractive applications based on the emergent phenomena at interfaces is a spin-orbit torque (SOT) device, in which the spin current generated by an in-plane charge current flowing along a spin channel exerts torques on a ferromagnetic (FM) layer, resulting in SOT-driven magnetization switching.[5] While the conventional SOT device consists of the spin channel made of heavy metals such as W and the FM layer,[6] recent demonstrations have been conducted with the SOT devices consisting of 2D spin channels and 2D FMs with a form of van der Waals heterojunction, i.e., topological semimetallic $WTe_2$/ferromagnetic $Fe_3GeTe_2$.[7,8] There are two advantages for the 2D spin channels, which are the high spin injection efficiency due to the atomically flat interface, and the moderate change of electric potential at the interface due to the same Te termination, which causes the reduction of spin loss across the $Fe_3GeTe_2$ and $WTe_2$ layers.[9] Furthermore, the 2D-$Fe_3GeTe_2$/Pt bilayer also exhibits the SOT-driven magnetization switching, although the interface does not form a van der Waals heterojunction.[10] Therefore, the 2D material has been promising for highly efficient SOT devices through interface engineering in the future.

MXene has recently become known as a new class of 2D materials.[11] The chemical formula is $M_{n+1}X_nT_x$, where the sites *M*, *X*, and *T* represent transition metals such as Ti and Cr, 2*p* light elements such as C and N, and surface terminations such as O and Cl, respectively. Specifically, *n* corresponds to the number of *M-X-M* bonds, which varies from 1 to 4, and *x* is a variable. These parameters can effectively modulate the physical and chemical properties of MXene,[12] because of the significant orbital hybridization between the elements *X* and *M*, which originates from the high electronegativity of the 2*p* light element *X*.[13] Thus, various applications have been demonstrated with the MXene, which are biomedicine,[14] mechanical science,[15] optoelectronics,[16] energy storage.[17]

As mentioned above, emergent phenomena at interfaces are of increasing interest in 2D materials, however, it is not yet clear what phenomena occur at the 2D-MXene/FM interface. As one of the examples, it is reported that the magnetic moment of the non-magnetic Cu and/or Pt is induced by the adjacent FMs, and the origin is explained by both magnetic coupling and charge transfer.[18,19] If the magnetic moment of Cr in the MXene layer is induced by the adjacent FM layer, which would have a great impact on the spin injection efficiency for the spintronic devices with 2D systems.[7,8] Therefore, the investigation is expected to strengthen the framework of interfacial engineering of 2D-MXene, leading to the significant approach for the efficient SOT-driven magnetization switching.

In this study, we aim to analyze the induced magnetic moment and the electronic states of Cr in the $Cr_2N$-MXene/Co bilayer, by means of both the X-ray magnetic circular dichroism (XMCD) and the angle-dependent hard X-ray photoemission spectroscopy (HAXPES). As a result, the magnetic moment of Cr was induced by Co, and the surface-sensitive HAXPES spectra of Cr 2*p* core level was modulated by Co as well. The combined study with XMCD and HAXPES identified that the exchange magnetic coupling between Cr and Co is one of the major origins for the induced magnetic moment of Cr in the 2D MXene, rather than the charge transfer due to different work functions.

## 2. Experimental details

### 2.1. Film preparation and characterization

The $Cr_2N$ 2D-MXene was deposited on the *c*-plane oriented $Al_2O_3$ substrate using the DC magnetron reactive nitridation sputtering with the Cr sputtering target at substrate temperature ($T_{sub}$) from room temperature (RT) to 650 °C, where the Nitrogen flow ratio defined as $N_2/(Ar + N_2)$ was varied from 2% to 25% to form stoichiometric $Cr_2N$ layer. The Co, Cu, and Al layers were deposited via DC magnetron sputtering at RT. The crystal structure was investigated via X-ray diffraction (XRD; SmartLab; Rigaku Corporation) with Cu-$K_\alpha$ radiation.

### 2.2. X-ray magnetic circular dichroism (XMCD) experiments

The XMCD measurements were performed at the BL14U Synchrotron Radiation Facility, NanoTerasu. Soft X-ray absorption spectra (XAS) were recorded using the total electron yield (TEY) method while scanning photon energy at RT. The XMCD signal was obtained by subtracting each XAS signal for circularly polarized light with positive and negative helicities. For the measurements of Cr and N, the XAS for each helicity was repeated five times and averaged to boost the signal-to-noise ratio. The magnetic field was applied perpendicularly to the surface of the sample.

### 2.3. X-ray photoemission spectroscopy (XPS) experiments

The conventional XPS of Al 2*s* core level was measured by the NIMS in-house system with the hemispherical analyzer (SES100, Scienta Omicron) and the excitation source of Al $K_\alpha$ without monochromator (DSX400, Scienta Omicron) at RT. The take-off angle (ToA) was varied at 90°, 60°, 30°, and 15° relative to the in-plane direction to adjust surface sensitivity. The data acquired at ToA = 90° corresponds to the normal emission configuration, which provides the highest bulk sensitivity configuration under the present experimental conditions.

*2.4. Hard X-ray photoemission spectroscopy (HAXPES) experiments*

The HAXPES experiment was performed at the BL09U Synchrotron Radiation Facility, NanoTerasu. The synchrotron X-ray with a photon energy of ~6 keV was irradiated to the film with two different ToAs of 10° and 88° with respect to the in-plane direction, that is, the photoemission from surface is dominant for ToA = 10°, comparing to that for ToA = 88°. Using a monolithic Woltermirror, the incident X-ray beam was focused to 7 × 10 (vertical × horizontal) µm$^2$. For the ToA = 88° condition, the footprint of the X-ray beam on the sample at grazing incidence was as small as 7 × 300 (vertical × horizontal) µm$^2$. The photoelectrons were detected and analyzed using a high-resolution hemispherical electron analyzer (R4000, Scienta Omicron). All measurements were performed at RT. The high-brilliance synchrotron radiation allowed us to detect tiny difference in spectra, which is influenced by the adjacent layer. The energy axis of HAXPES spectra were calibrated based on the binding energy ($E_b$) and the peak positions of Au $4f_{7/2}$ and Fermi level ($E_F$) of Au.

**3. Results and discussion**

*3.1. Crystal structure*

Figure 1(a1) shows the unit cell of Cr$_2$N MXene with a hexagonal structure, of which lattice constants are $a = b = 0.48$ nm and $c = 0.45$ nm. The collinear antiferromagnetic structure of Cr has been reported for a wide temperature range from 100 K to 500 K.[20] Figures 1(a2) and 1(a3) depict the supercell model of Cr$_2$N ($1\bar{1}00$) plane and (0001) plane, respectively. Figure 1(b) shows the out-of-plane XRD profiles for the 15-nm-thick Cr$_2$N single layer depending on the N$_2$ flow ratio, while reactive nitridation sputtering of Cr. Two XRD peaks appeared at $2\theta/\omega \approx 42°$ and 44° for $Q = 2\%$, suggesting coexisting of pure Cr and $Cr_2N$ phases. The XRD peak appeared at $2\theta/\omega \approx 40°$ corresponds to the $Cr_2N$ (0001) plane, suggesting no coexisting phases for $Q = 15\%$. Two XRD peaks appeared at $2\theta/\omega \approx 37°$ and $40°$ for the $Q$ higher than 20%, which correspond to the CrN (111) and $Cr_2N$ (0001) planes, respectively. Therefore, we determined the optimum $Q$ for the single $Cr_2N$ MXene phase as $Q = 15\%$. Figures 1(c1) and 1(c2) show the XRD pole figure and $\varphi$-scan profile for the $Cr_2N$ ($\bar{1}\bar{1}21$), suggesting the hexagonal structure with six-fold in-plane crystal symmetry in the $Cr_2N$ film.

### 3.2. XMCD to study induced magnetic moment of Cr

We measured the element-selective magnetic properties at the $Cr_2N$-MXene/FM interfaces by means of XMCD for two samples: (a) $Cr_2N$(5 nm)/Co(1 nm), and (b) $Cr_2N$(5 nm)/Pt(1 nm) [Figs. 2(a) and 2(b)]. The X-ray absorption spectra (XAS) near the $L_{2,3}$-edge of Cr exhibited two peaks. The XMCD signal was evident near the Cr $L_2$-edge for the $Cr_2N$/Co, while it was not observed for the $Cr_2N$/Pt. Using the sum rule (see Fig. S1 in the Supporting Information) for the results of $Cr_2N$/Co bilayer, spin ($m_{spin}$) and orbital ($m_{orb}$) magnetic moments of Cr were estimated to be $-0.063$ $\mu_B$ and $\sim 0$ $\mu_B$, respectively. The magnitude of induced moment of Cr is 25 times smaller than that of the calculated value of Co ($m_{spin} \approx 1.63$ $\mu_B$ and $m_{orb} \approx 0.1$ $\mu_B$).[21] The $m_{spin}$ of Cr corresponds to the uncompensated moment ($m_{Cr}^{UC.}$) originating from the imbalance in the antiferromagnetic structure of $Cr_2N$ due to the adjacent Co layer.

### 3.3. XPS to study surface oxidation state

To investigate the surface oxidation state of the 1-nm-thick Al film used as a capping layer, angle-dependent XPS was employed to analyze the Al $2s$ core level. Figures 3(a1-a4) show the XPS spectra for the $Cr_2N$(5 nm)/Al(1 nm) with various ToA. Notably, spectra acquired lower ToA detects more photoelectrons from the top surface. The spectra were reproduced with two peaks, namely, the peak 1 and peak 2 that are originating from the metallic Al and Al-O, respectively. Figure 3(b) shows the ToA dependences of $E_b$ and intensity ratio ($I_1/I_2$), where the $I_{1(2)}$ denotes integral intensity of the peak 1(2). $E_b$ for the peak 1 slightly decreased with increasing ToA, which was consistent with that for the peak 2. In addition, the $I_1/I_2$ increased monotonically from ~0.32 to ~0.73 with increasing ToA, indicating that the metallic Al component is more dominant at higher ToAs compared to Al-O, which is more pronounced at lower ToA. The superimposed curve was obtained from calculation based on Ref. [22]. The inelastic mean free path was derived from Ref. [23], assuming inorganic materials. The calculated result closely reproduced the experimental result, yielding a surface oxide thickness of ~0.2 nm. We thus conclude that the Al-O was present only near the surface of the Al capping layer, which was estimated to be ~0.2 nm. It is inferred that the natural oxidation of the $Cr_2N$ surface can be ruled out from the consideration of HAXPES spectra show below.

### *3.4. HAXPES to study electronic state of Cr*

Figure 4(a) shows the film stacking structure to evaluate the electronic states of Cr influenced by the adjacent Co [sample (i)], Cu [sample (ii)], and Al [sample (iii)]. The controlled sample (iv) acts as a reference. These samples were prepared for the following purposes. First, comparison between the samples (i) and (ii) allows for separating the possible origins of $m_{Cr}^{UC.}$, such as the magnetic coupling effect by the adjacent Co layer or the charge transfer by different work function (WF), because the WF of Co is ~5.0, which is consistent with that of (111) plane-oriented Cu.[24] Namely, the origin of $m_{Cr}^{UC.}$ can be determined as the magnetic coupling effect, when no change in HAXPES spectra were observed between the samples (i) and (ii). Second, the sample (iii) provides the electronic states of Cr in the pristine $Cr_2N$ MXene.

Figure 4(b1) shows the HAXPES spectra for the sample (i) with the ToA of 10° and 88°. Spectral normalization was performed outside the effective energy range, on both the high-binding energy and low-binding energy sides. The subsequent spectra in Figs. 4(b2)-4(e) were normalized using the same procedure. The peaks appeared at the $E_b$ of 584 eV and 574 eV, which correspond to the Cr $2p_{1/2}$ and $2p_{3/2}$ core levels, respectively. Each peak exhibited two components: a main peak and a satellite at a higher binding energy, as indicated by the arrows. With the present normalization, the intensity of the main peak at ToA = 88° was larger than that at ToA = 10°. In contrast, the satellite was more prominent at ToA = 10°. These observations suggest that the main peak reflects the characteristics of the entire film, whereas the satellite is associated with the interface of sample (i). Note that the intensity variation of the satellite, as indicated by arrows, suggests that the electronic state of Cr was modulated by the adjacent Co layer. Figures 4(b2) and 4(b3) show corresponding results for the sample (i) for the N 1$s$ and O 1$s$ core levels, respectively. The hump-like feature appeared at lower $E_b$ than 1$s$ main peak at ToA = 10°, whereas no such feature was observed at ToA = 88°. This can be attributed to the bonding state of N with Co at the $Cr_2N$/Co interface. The intensity of O 1$s$ core level was much higher at ToA = 10°, comparing to the case of ToA = 88°. This result is consistent with the surface oxidation of the Al capping layer, as evaluated by conventional XPS in Section 3.3.

To investigate the effect of the magnetic moment in Co layer to the $Cr_2N$ layer, the same measurements were conducted on the sample (ii), in which the $Cr_2N$ layer has an interface with a Cu layer instead of a Co layer. Figure 4(c) presents the Cr 2$p$ core level spectra of sample (ii). We can see the same results, that is, a prominent main peak appeared at ToA = 88°, while the satellite is more prominent at ToA = 10°. Thus, the electronic state of Cr was modulated by Cu layer, resulting from the bonding state between Cr and Cu. Comparing the spectra of the samples (i), (ii), and (iii) for ToA = 10°, it was revealed that the electronic state of Cr near the interface remained consistent regardless of whether the adjacent layer was Co or Cu, as shown in Fig. 4(d). The spectra for the sample (iii) shows smaller intensity at the $E_b$ from 573 eV to 590 eV. On the other hand, the discrepancy becomes negligible for the sample (i), (ii), and (iii) with ToA = 88°, as shown in Fig. 4(e), indicating that the discrepancy is confined to the interface. One possible explanation is the difference of the WFs of Co, Cu and Al. Co and Cu have similar WFs (WF ≈ 5.0), whereas Al has a lower one (WF ≈ 4.1). At the interface between materials with different WF, charge redistribution may occur. This effect is expected to be the similar for Co and Cu due to their nearly identical WFs, whereas it differs in the case of Al.

In discussion, we consider the observed XMCD signal as shown in Fig. 2. The induced magnetic moment of Cr, $m_{Cr}^{UC.}$, was evident by the adjacent Co layer at the Cr$_2$N/Co interface, while it was not observed at the Cr$_2$N/Pt interface, in which the WF of Co (~5.0) is smaller than that of Pt (~5.7). To exclude the influence from the different WF, we replaced the Pt with Cu, of which WF is similar to that of Co. As shown in Fig. 4(d), the interfacial electronic state of Cr$_2$N/Co is consistent with that of Cr$_2$N/Cu, which confirms that the interlayer magnetic coupling is one of the major origins for the $m_{Cr}^{UC.}$ in the Cr$_2$N 2D-MXene, rather than the charge redistribution. Within the framework of 2D ferromagnetism, van der Waals ferromagnetic 2D materials have recently attracted considerable attention due to their potential for practical applications. For example, Fe$_3$GaTe$_2$ exhibits intrinsic ferromagnetism with Curie temperature above RT and sizable perpendicular magnetic anisotropy.[25] This is mostly originating from the electronic state of 3$d$ orbitals of transition metals that is modulated by the various lattice symmetry such as honeycomb and triangle structures, which is classified as the intrinsic ferromagnetism of pristine 2D materials. On the other hand, some reports show emergent and/or tailored 2D magnetism by pressure and elemental doping,[26,27] which is classified as the extrinsic one. In addition, magnetic proximity effect has been examined in 2D van der Waals heterojunction systems.[28,29] Although various mechanisms are reported mentioned above, emergent ferromagnetism due to surface termination is a unique characteristics for MXene, e.g., F- and OH-terminated $Cr_2C$ and/or $Cr_2N$ are predicted to be ferromagnetic.[30] Therefore, the findings in this study that the magnetic exchange coupling in the MXene/FM system could open another pathway to induce the ferromagnetism in the MXene, and leading to a phenomena of spin-filtering effect in the 2D devices.[31]

## 4. Conclusion

To examine the origin of induced magnetic moment of Cr, $m_{Cr}^{UC.}$, in the $Cr_2N$ MXene/Co bilayer, we evaluated interfacial electronic states via angle-resolved HAXPES at RT with sufficient sensitivity at NanoTerasu synchrotron radiation facility. The controlled $Cr_2N$ MXene/Cu was prepared for comparison, which of these samples allow us to distinguish the possible major origins, that is, the interfacial magnetic coupling and the charge transfer. The interface sensitive HAXPES spectra for the $Cr_2N$/Co was consistent with that for the $Cr_2N$/Cu. Furthermore, bulk sensitive ones for both samples were consistent. These results led us to conclude that the inter-layer magnetic coupling can be a major origin for the induced magnetic moment in the 2D MXene adjacent to FM, rather than charge transfer due to the different WFs.

## Acknowledgments

The XMCD measurements were performed at the BL14U of the synchrotron radiation facility NanoTerasu. This work was supported by KAKENHI Grants-in-Aid No. 23K22803 from the Japan Society for the Promotion of Science (JSPS). Part of this work was performed under the Cooperative Research Project Program of the RIEC, Tohoku University.

**Figures**

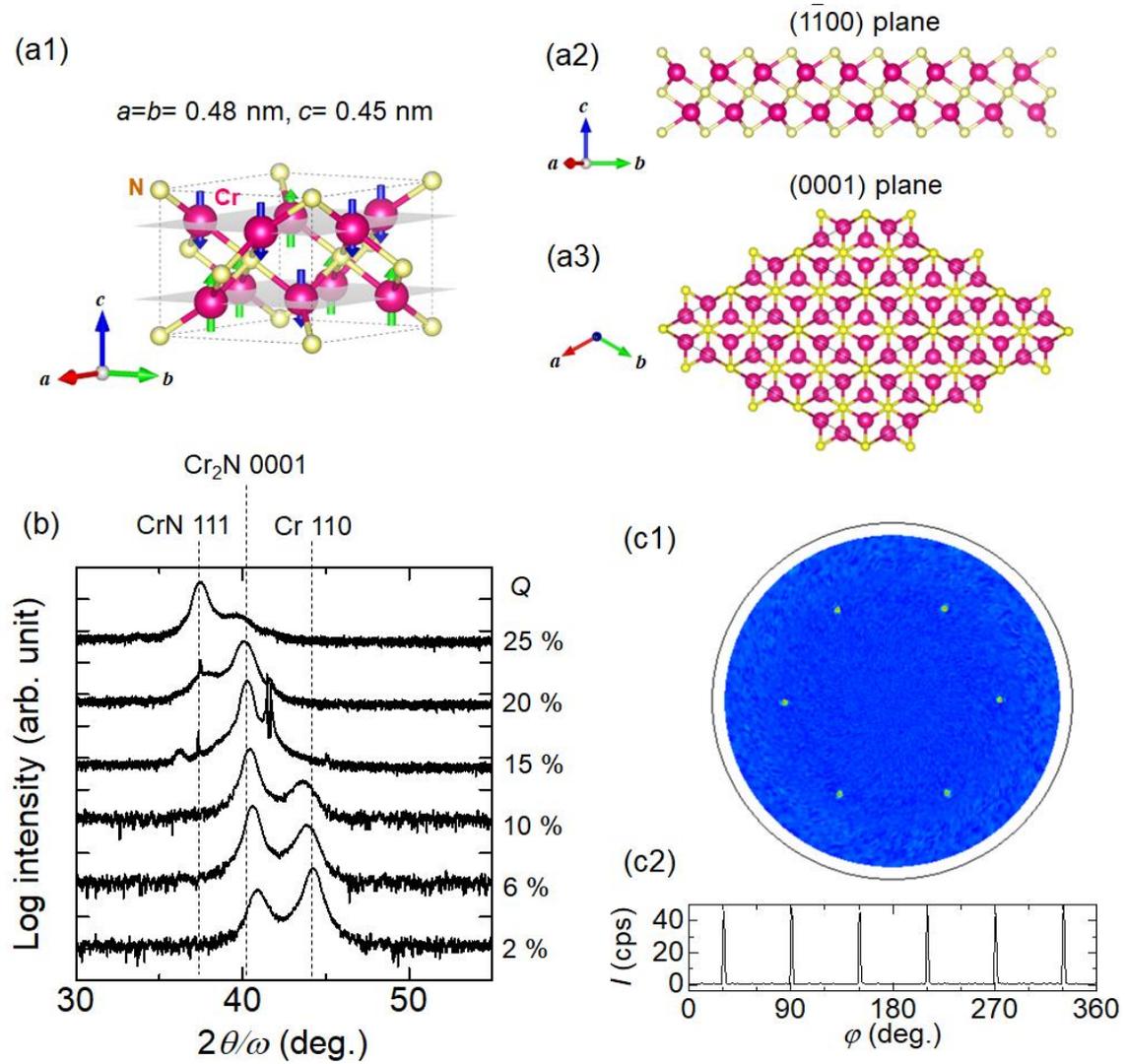

**Figure 1.** (a1-a3) Unit-cell model of $Cr_2N$ MXene together with possible magnetic structure (a1), cross-sectional plane view (a2) and top view (a3) of the $Cr_2N$ supercell. (b) Out-of-plane XRD profiles for 15-nm-thick $Cr_2N$ films with various Nitrogen ratio relative to Argon $Q = N_2/(Ar+N_2)$, while sputtering deposition on $c$-plane oriented $Al_2O_3$ substrates. (c1, c2) XRD pole figure (c1) and $\varphi$-scan profile (c2) for the $Cr_2N$ ($\bar{1}\bar{1}21$).

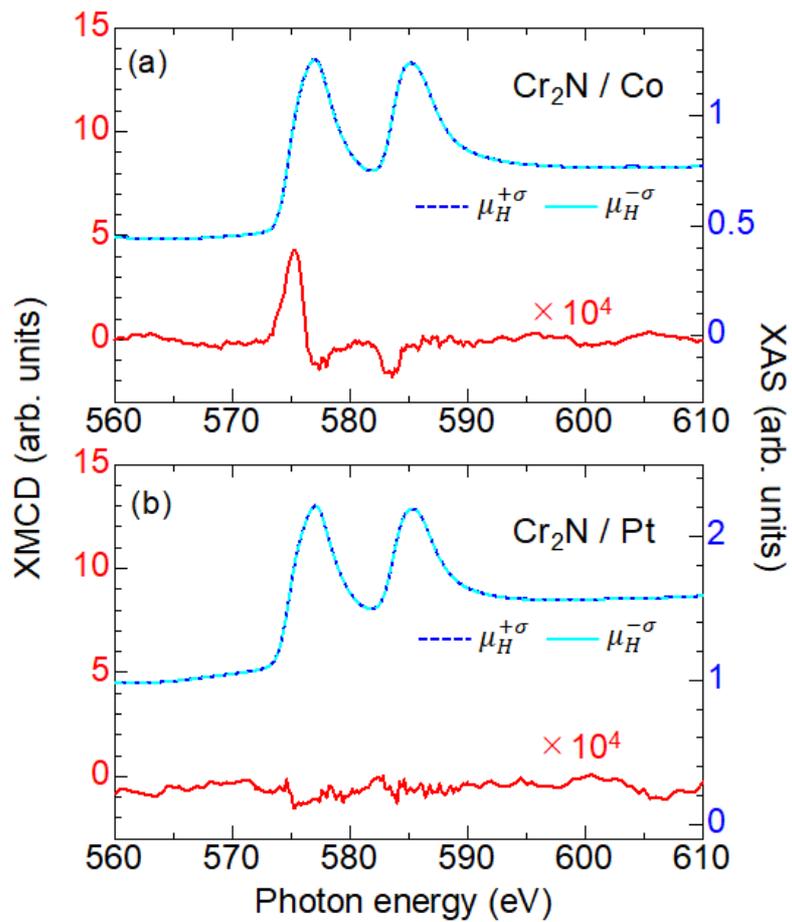

**Figure 2.** (a) XMCD (red) and XAS (blue) spectra for the Cr $L_{2,3}$-edge of the sample, $Al_2O_3$ sub.//$Cr_2N$(5 nm)/Co(1 nm)/Al(1 nm). (b) Same result as (a) but for the sample of $Cr_2N$(5 nm)/Pt(1 nm).

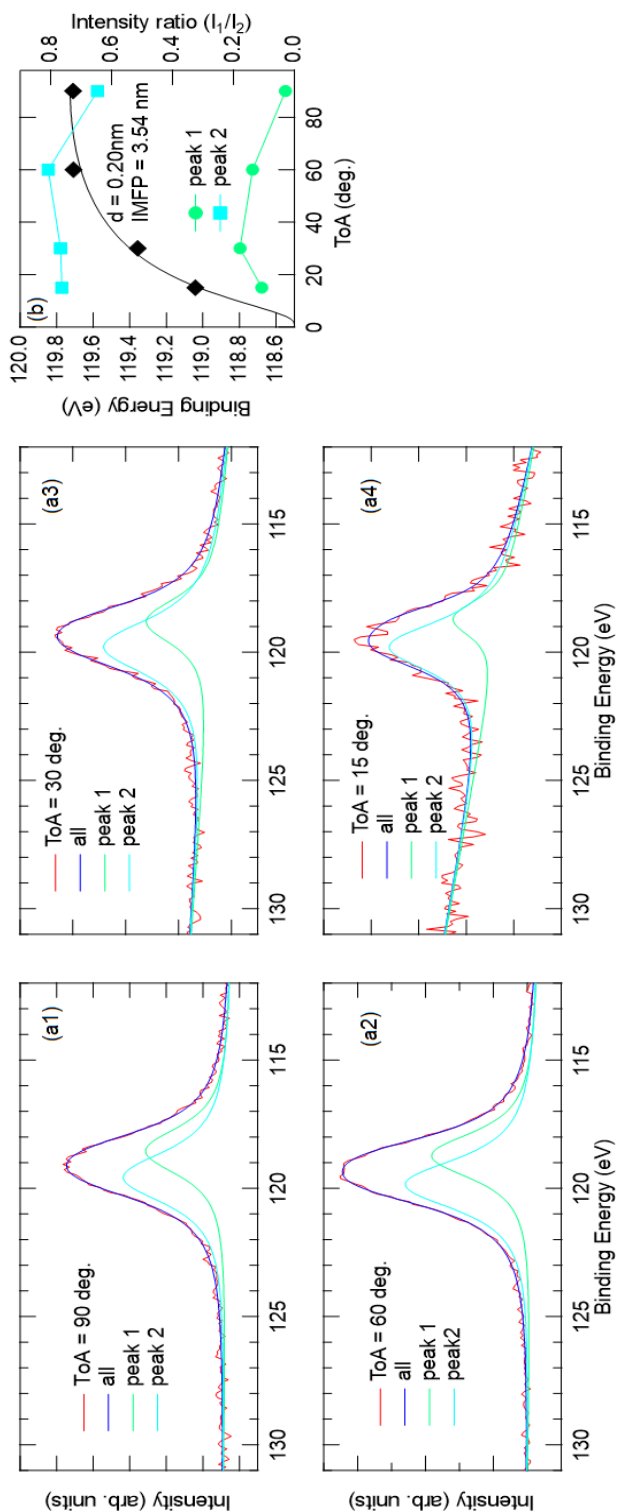

**Figure 3. (See Figs.ppt uploaded for enlarged one)** (a1-a4) XPS spectra of Al 2$s$ core level for the sample, Al$_2$O$_3$ sub.//Cr$_2$N(5 nm)/Al(1 nm), with ToA of 90°, 60°, 30°, and 15°, where the spectra were fitted by two peaks, which are the peak 1 and peak 2 originating from the metallic Al and Al-O, respectively. Note that the spectra with lower ToA detects more photoelectrons from top surface. (b) ToA dependences of binding energy and intensity ratio ($I_1/I_2$), where the $I_{1(2)}$ denotes integral intensity of the peak 1(2).

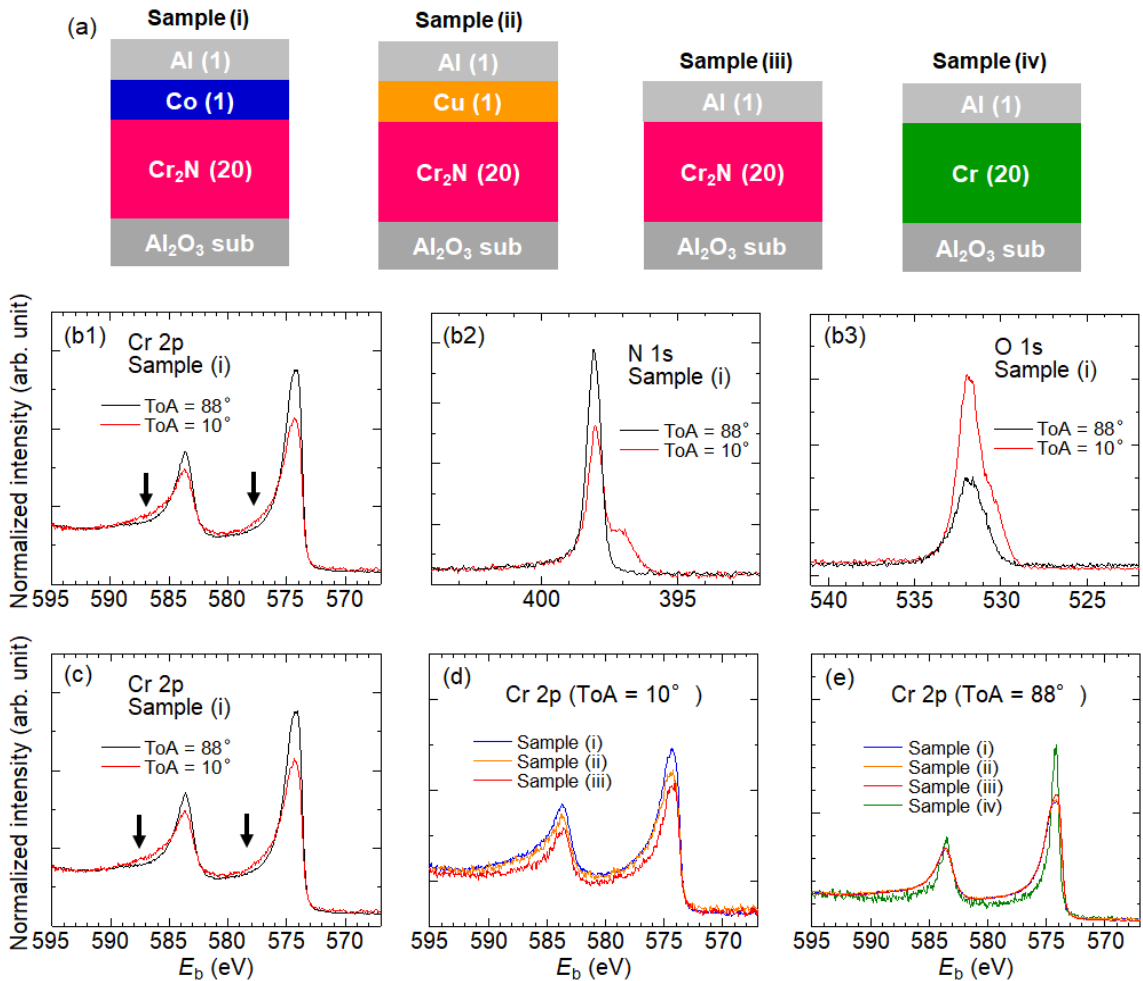

**Figure 4.** (a) Sample variations to investigate the electronic state of Cr affected by the adjacent Co [sample (i)], Cu [sample (ii)], and Al layers [sample (iii)]. The sample (iv) corresponds to pure Cr as a control sample. The number in parentheses represents the layer thickness. (b1-b3) HAXPES spectra for the Cr 2*p* core level (b1), N 1*s* core level (b2), and O 1*s* core level (b3) of the sample (i). Black and red lines correspond to the spectra with ToA = 88° and 10°, respectively. (c) Same measurements as (b1) but for the sample (ii). (d,e) Comparison among the spectra for ToA = 10° (d) and that for ToA = 88° (e).